\RequirePackage{lineno}
\documentclass[prl,twocolumn,amsmath,amssymb]{revtex4}
\usepackage{overpic}
\usepackage{graphicx}
\usepackage{epsfig}
\usepackage{dcolumn}
\usepackage{bm}
\usepackage{rotating}
\usepackage{multirow}
\usepackage{subfigure}
\usepackage{hhline}
\usepackage{amssymb}
\usepackage{lineno}
\usepackage{hyperref}
\hypersetup{colorlinks,linkcolor=blue,anchorcolor=blue,citecolor=blue}

\newcommand{\EE}{e^+e^-}

\newcommand{\jpsi}{J/\psi}

\newcommand{\bfg}{\begin{figure}}
\newcommand{\efg}{\end{figure}}
\newcommand{\bitm}{\begin{itemize}}
\newcommand{\eitm}{\end{itemize}}
\newcommand{\bnum}{\begin{enumerate}}
\newcommand{\enum}{\end{enumerate}}
\newcommand{\btbl}{\begin{table}}
\newcommand{\etbl}{\end{table}}
\newcommand{\btbu}{\begin{tabular}}
\newcommand{\etbu}{\end{tabular}}
\newcommand{\bcl}{\begin{center}}
\newcommand{\ecl}{\end{center}}

\newcommand{\beq}{\begin{equation}}
\newcommand{\eeq}{\end{equation}}
\newcommand{\beqr}{\begin{eqnarray}}
\newcommand{\eeqr}{\end{eqnarray}}

\newcommand{\blue}{\color{blue}}

\newcommand{\br}{{\cal B}}

\renewcommand{\EE}{e^+e^-}

\newcommand{\psipp}{\psi(3770)}
\renewcommand{\jpsi}{J/\psi}
\newcommand{\mev}{\mathrm{MeV}}

\newcommand{\gev}{\mathrm{GeV}}
\newcommand{\gevc}{\mathrm{GeV}/c}
\newcommand{\gevcc}{\mathrm{GeV}/c^2}
\newcommand{\invfb}{\mathrm{fb}^{-1}}

\newcommand{\pb}{\mathrm{pb}}
\newcommand{\bw}{\mathrm{BW}}

\let\oldequation\equation
\let\oldendequation\endequation

\renewenvironment{equation}
  {\linenomathNonumbers\oldequation}
  {\oldendequation\endlinenomath}

\begin{document}

\title{\boldmath Observation of $\psi(3770)\to\eta J/\psi$ }

\author{
\begin{small}
\begin{center}
M.~Ablikim$^{1}$, M.~N.~Achasov$^{10,b}$, P.~Adlarson$^{69}$, S. ~Ahmed$^{15}$, M.~Albrecht$^{4}$, R.~Aliberti$^{29}$, A.~Amoroso$^{68A,68C}$, M.~R.~An$^{33}$, Q.~An$^{65,51}$, X.~H.~Bai$^{59}$
% ...
\\
\vspace{0.2cm}
(BESIII Collaboration)\\
\vspace{0.2cm} {\it
$^{1}$ Institute of High Energy Physics, Beijing 100049, People's Republic of China\\
% ...
$^{4}$ Bochum Ruhr-University, D-44780 Bochum, Germany\\
% ...
$^{10}$ G.I. Budker Institute of Nuclear Physics SB RAS (BINP), Novosibirsk 630090, Russia\\
% ...
$^{15}$ Helmholtz Institute Mainz, Staudinger Weg 18, D-55099 Mainz, Germany\\
% ...
$^{29}$ Johannes Gutenberg University of Mainz, Johann-Joachim-Becher-Weg 45, D-55099 Mainz, Germany\\
% ...
$^{33}$ Liaoning Normal University, Dalian 116029, People's Republic of China\\
% ...
$^{51}$ State Key Laboratory of Particle Detection and Electronics, Beijing 100049, Hefei 230026, People's Republic of China\\
% ...
$^{59}$ University of Jinan, Jinan 250022, People's Republic of China\\
% ...
$^{65}$ University of Science and Technology of China, Hefei 230026, People's Republic of China\\
% ...
$^{68}$ University of Turin and INFN, (A)University of Turin, I-10125, Turin, Italy; (B)University of Eastern Piedmont, I-15121, Alessandria, Italy; (C)INFN, I-10125, Turin, Italy\\
$^{69}$ Uppsala University, Box 516, SE-75120 Uppsala, Sweden\\
% ...
$^{b}$ Also at the Novosibirsk State University, Novosibirsk, 630090, Russia\\
% ...
}
\end{center}
\vspace{0.4cm}
\end{small}
 }
\noaffiliation{}

\date{\today}

\begin{abstract}
The Born cross section of the process $e^+e^-\to\eta J/\psi$ at a center-of-mass 
energy $\sqrt{s}=3.773~\gev$ is measured to be ($8.89\pm0.88\pm0.42$)~$\pb$,
using a data sample collected with the BESIII detector operating at the BEPCII 
storage ring. The decay $\psi(3770)\rightarrow\eta J/\psi$ is observed 
for the first time with a statistical significance of $7.4\sigma$. From a fit to the 
dressed cross-section line-shape of $\EE\to\eta\jpsi$ from $\sqrt{s}=3.773$ to $4.600~\gev$
we obtain the branching fraction of the decay $\psi(3770)\to \eta J/\psi$ to be $(11.6\pm6.1\pm1.0)\times 10^{-4}$ when the $\psi(3770)$ decay amplitude is added coherently to the other contributions, and
$(7.9\pm1.0\pm0.7)\times 10^{-4}$ when it is added incoherently. Here the first uncertainties are statistical and the second  
are systematic.
\end{abstract}

\maketitle
\oddsidemargin -0.2cm
\evensidemargin -0.2cm

The nature of the $\psi(3770)$ resonance is still a subject of debate. 
Conventionally, the $\psi(3770)$  has been regarded as the lowest-mass $D$-wave charmonium 
state above the $D\bar{D}$ threshold, {\it i.e.} a pure $c\bar{c}$ meson in the quark 
model~\cite{Eichten:1978tg}.  However, in Ref.~\cite{Voloshin:2005sd} it is suggested that 
the $\psi(3770)$ may contain a considerable tetra-quark component. This would help to 
explain several unsolved issues of charmonium physics that are at variance with the standard 
theoretical expectations, namely, the large non-$D\bar{D}$ decay width of the state, the 
abnormal ratio of the branching fractions of $\psi(3770) \to D^+D^-$ and $\psi(3770) \to D^0 \bar{D}^0$,
and the $\rho-\pi$ puzzle~\cite{Mo:2006cy}. 
Additionally, a large tetra-quark component would suppress the re-annihilation 
mechanism in the charmless final states of $\psi(3770)$ decays,
which will enhance penguin amplitudes (particularly in $b \to s$ transitions)
in $B$ decays and non-$K\bar{K}$ decays of $\phi$ meson~\cite{Rosner:2004wy}.
Recently, BESIII reported evidence for $e^+ e^- \to \pi^+ \pi^- \psi(3770)$ at the 
center-of-mass (c.m.) energies $\sqrt{s}=4.26$ and $4.36~\gev$~\cite{BESIII:2019tdo}, 
indicating a possible link between the $\psi(3770)$ and the tetra-quark candidates $Y(4260)$ and $Y(4360)$~\cite{Maiani:2005pe,Chen:2016qju,Olsen:2017bmm,Brambilla:2019esw,Zhu:2021vtd}.  
Therefore, improved knowledge of the nature of the $\psi(3770)$ will also deepen the understanding 
of the nature of exotic charmonium-like (also called $\rm XYZ$) states and more generally
the non-perturbative behavior of the strong interaction.

As suggested in Ref.~\cite{Voloshin:2005sd}, a large tetra-quark component in the 
$\psi(3770)$ would lead to an enhancement of the hadronic transition 
$\psi(3770)\to \eta J/\psi$, with a predicted branching fraction of $\sim 15 \times 10^{-4}$. 
Previously, CLEO studied this decay under the assumption of no interference between 
the resonant decay and the continuum process. The branching fraction is measured to be
$\br(\psi(3770)\to\eta\jpsi)=(8.7 \pm 3.3 \pm 2.2)\times 10^{-4}$, compatible 
with the prediction of Ref.~\cite{Voloshin:2005sd}, and corresponds to a signal with a 
statistical significance of $3.5\sigma$~\cite{CLEO:2005zky}. Usually, the branching 
fraction of $\psi(3770)\to \eta J/\psi$ is utilized as an input in theoretical calculations 
of the hadronic transition properties of excited charmonium or charmonium-like states,
both in the large-distance meson loop~\cite{Zhang:2009kr,Li:2013zcr} and the Nambu–Jona-Lasinio model~\cite{Anwar:2016mxo}.
In order to advance our understanding, it is desirable to obtain a more precise measurement with proper consideration of the possible interference between the resonant decay and the continuum process.

In this Letter, we report the measurement of the Born cross 
section of $e^+e^-\to\eta J/\psi$  using $2.93~\invfb$ of data~\cite{BESIII:2018dpx} taken at 
$\sqrt s=3.773$ GeV with the BESIII detector. The branching fraction of 
$\psi(3770)\to\eta J/\psi$ is determined by fitting the dressed cross-section line-shape with a  
combination of previous measurements~\cite{BESIII:2020bgb}. The $ J/\psi $ 
is only reconstructed through its decay to di-muons, while the di-electron decay is not used 
because of the high contamination from the radiative Bhabha process.

The BESIII detector is a magnetic spectrometer~\cite{BESIII:2009fln} located 
at the Beijing Electron Positron Collider (BEPCII). The cylindrical core of 
the BESIII detector consists of a helium-based multilayer drift chamber (MDC), 
a plastic scintillator time-of-flight system (TOF), and a CsI (Tl) 
electromagnetic calorimeter (EMC), which are all enclosed in a superconducting
solenoidal magnet providing a $1.0$~T magnetic field. The solenoid is supported 
by an octagonal flux-return yoke with resistive plate counter muon identifier 
modules interleaved with steel. The acceptance of charged particles and photons 
is $93$\% over $4\pi$ solid angle. The charged-particle momentum resolution at
$1~\gevc$ is $0.5\%$, and the specific ionization energy loss resolution is 
$6\%$ for the electrons from Bhabha scattering. The EMC measures photon energies 
with a resolution of $2.5\%$ ($5\%$) at $1~\gev$ in the barrel (end-cap) region. 
The time resolution of the TOF barrel part is $68~\mathrm{ps}$, while that of 
the end-cap part is $110~\mathrm{ps}$.

Large samples of simulated events produced with the \textsc{GEANT4} based~\cite{GEANT4:2002zbu} 
Monte Carlo (MC) software, which includes the geometric description of the BESIII
detector and the detector response, are used to determine the detection efficiency 
and to estimate the background contribution. The simulation includes the beam-energy
spread and initial-state radiation (ISR) in the $e^+e^-$ annihilations modeled 
with the generator \textsc{KKMC}~\cite{Jadach:2000ir,Jadach:1999vf}. 
The decays $\psi(3770)\rightarrow\eta J/\psi$, $J/\psi \rightarrow \mu^+\mu^-$, 
and $\eta\to\gamma\gamma$ are generated with the \textsc{VVS} (Vector Vector Scalar)
P\_wave, \textsc{VL} (Vector Lepton Lepton), and phase space
(\textsc{PHSP}) configurations of
\textsc{EVTGEN}~\cite{Lange:2001uf,Ping:2008zz}, respectively. The
inclusive MC samples consist  
of the production of the $D\bar{D}$ pairs, the non-$D\bar D$ decays of the
$\psi(3770)$, the ISR production of $J/\psi$ and $\psi(3686)$ states, and 
the continuum processes ($e^+e^-\rightarrow u\bar{u}, d\bar{d}, s\bar{s}$) incorporated 
in \textsc{KKMC}~\cite{Jadach:2000ir,Jadach:1999vf}. The known decay modes are 
modeled with \textsc{EVTGEN}~\cite{Lange:2001uf,Ping:2008zz} using branching 
fractions taken from the Particle Data Group (PDG)~\cite{ParticleDataGroup:2020ssz},
and the remaining unknown decays from the charmonium states with
\textsc{LUNDCHARM}~\cite{Chen:2000tv,Yang:2014vra}. 
The final-state radiation from charged particles is incorporated with the 
\textsc{PHOTOS} package~\cite{Barberio:1993qi}.

Each candidate event is required to have two charged tracks with zero net charge, 
and at least two photon candidates. For each charged track, the distance of the 
closest approach to the interaction point is required to be less than $1~\mathrm{cm}$
in the radial direction and less than $10~\mathrm{cm}$ along the beam axis. The polar 
angle $\theta$ of the tracks with respect to the axis of the MDC must be within the fiducial volume of the MDC 
($|\cos\theta| < 0.93$). Photon candidates are reconstructed from isolated showers 
in the EMC which are at least $10^\circ$ away from the nearest charged track. The 
photon energy is required to be at least $25~\mev$ in the barrel region
($|\cos\theta|<0.80$) or $50~\mev$ in the end-cap region ($0.86<|\cos\theta|<0.92$).
In order to suppress electronic noise and energy depositions which are unrelated 
to the event, the difference between the EMC time and the event start time is 
required to satisfy $0\leq t \leq700~\mathrm{ns}$.

Tracks with momentum greater than $1~\gevc$ and energy deposited in the EMC less
 than $0.4~\gev$ are assumed to be muon candidates from $\jpsi$ decay. 
A vertex fit is performed for the two charged tracks, constraining them to originate from 
the interaction point. To improve the resolution and suppress
 background, a four-constraint (4C) kinematic fit is applied for the candidate events, imposing energy-momentum
conservation under the hypothesis of $e^+e^-\rightarrow\gamma\gamma\mu^+\mu^-$. 
In the case the event has  more than two photon 
candidates, all photon pairs are tested in the kinematic fit and the combination 
with the smallest value of  $\chi^2_{{\rm 4C}}$ is retained. The events are required 
to satisfy $\chi^2_{{\rm 4C}}<48$ to be retained for further analysis. This requirement is 
set by optimizing a figure-of-merit, defined as $\frac{S}{\sqrt{S+B}}$, 
where $S$ is the number of signal MC events and $B$ is 
the number of background events from the inclusive MC samples. The values of $S$ and $B$ are
normalized according to the integrated  luminosity and the branching fraction of 
$\psipp\to\eta\jpsi$ from the CLEO measurement~\cite{CLEO:2005zky}. 
To further suppress background events, the higher and lower energy photons are
required to satisfy $E_{\gamma \rm high}<0.52~\gev$ and 
$E_{\gamma \rm low}>0.135~\gev$, respectively. To remove contamination from the background process
$\psi(3770)\rightarrow\gamma\chi_{c1},\chi_{c1}\rightarrow\gamma J/\psi,J/\psi\rightarrow\mu^+\mu^-$, 
any event with $0.239~\gev< E_{\gamma \rm low} < 0.259~\gev$ and 
$0.377~\gev< E_{\gamma \rm high} < 0.396~\gev$ is removed.

\begin{figure}[hbtp]
\bcl
\subfigure{
\begin{minipage}{0.45\textwidth}
\centering
\includegraphics[width=\textwidth]{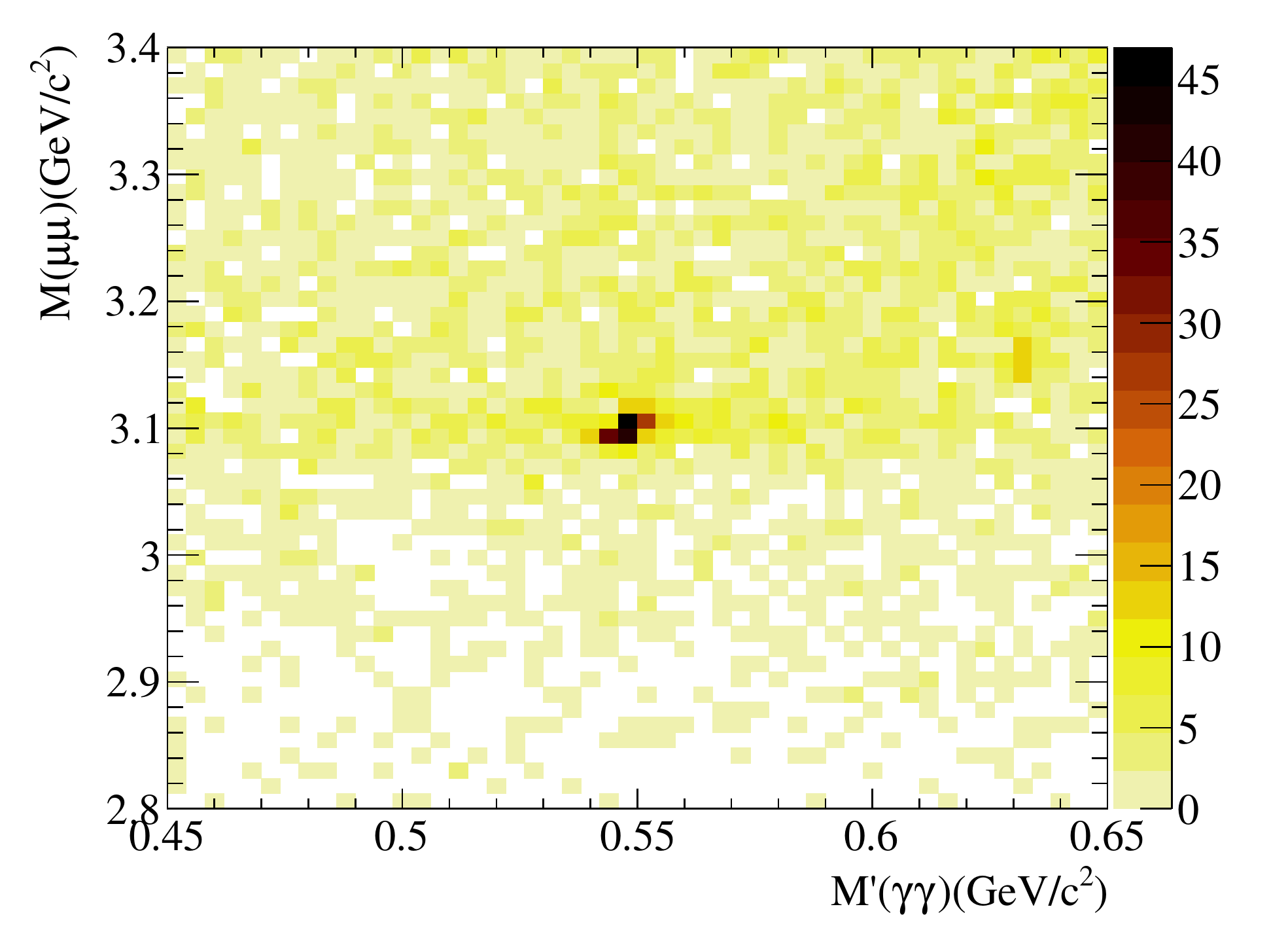}
\end{minipage}}
\caption{ Distribution of $M'(\gamma\gamma)$  versus $M(\mu\mu)$ of the  candidate events for $e^+e^-\to \eta J/\psi$ in data.}
\label{figure1}
\ecl
\end{figure}

Figure~\ref{figure1} presents the distribution of the modified invariant mass 
of the $\gamma\gamma$ pair ($M'(\gamma \gamma)$) against the  invariant mass of the $\mu^+\mu^-$ 
pair ($M(\mu\mu)$) for the  events in data after applying all the selection
criteria. Here $M'(\gamma \gamma) \equiv M(\gamma \gamma) + M(\mu \mu) - m_{J/\psi}$, where 
$m_{J/\psi}$ is the known $J/\psi$ mass~\cite{ParticleDataGroup:2020ssz}. A clear accumulation of signal is
observed around the intersection of the $J/\psi$ and $\eta$ mass regions.

\begin{figure*}[hbtp]
\bcl
\subfigure{
\begin{minipage}{0.3\textwidth}
\centering
\includegraphics[width=\textwidth]{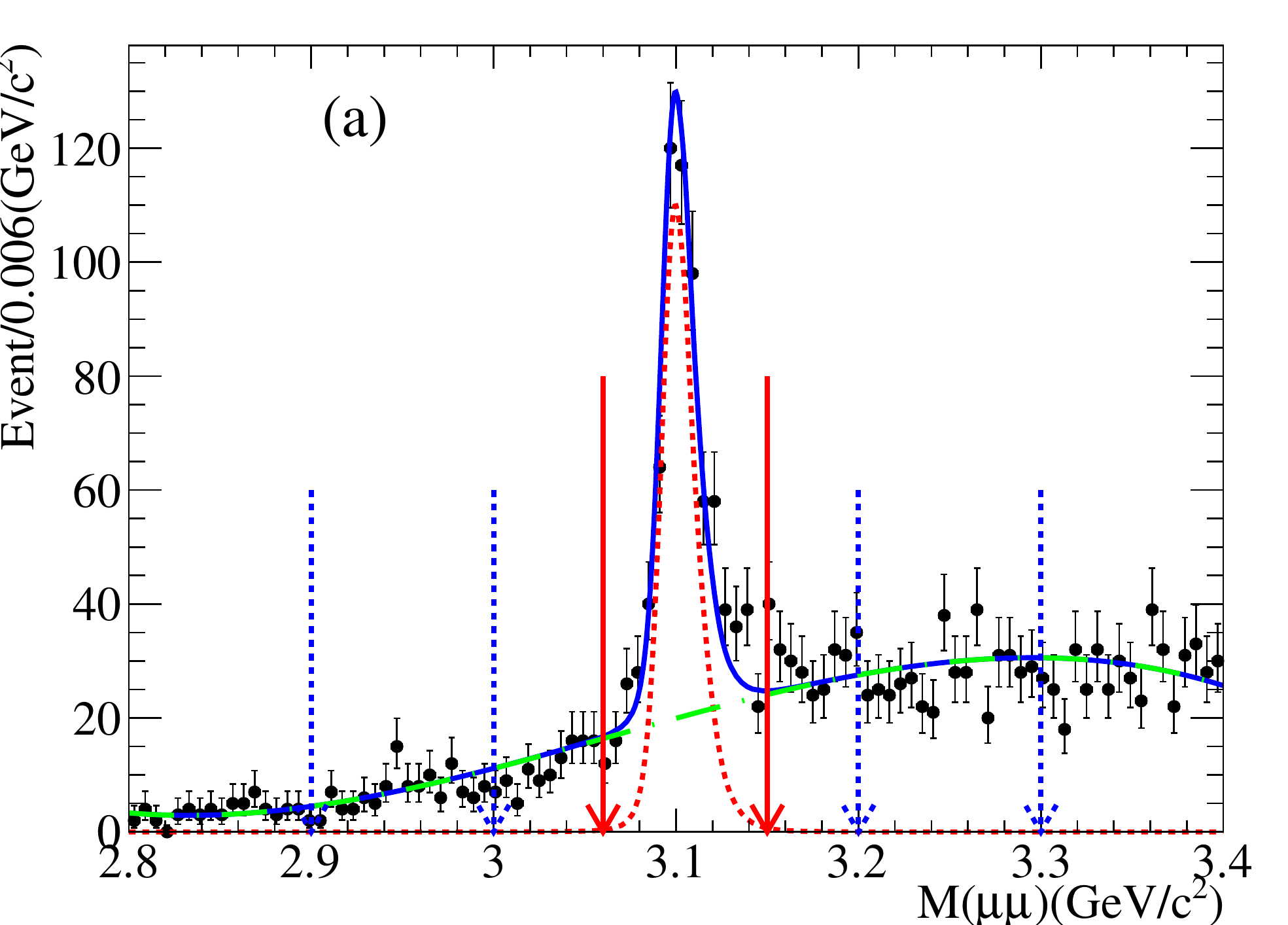}
\end{minipage}
\begin{minipage}{0.3\textwidth}
\centering
\includegraphics[width=\textwidth]{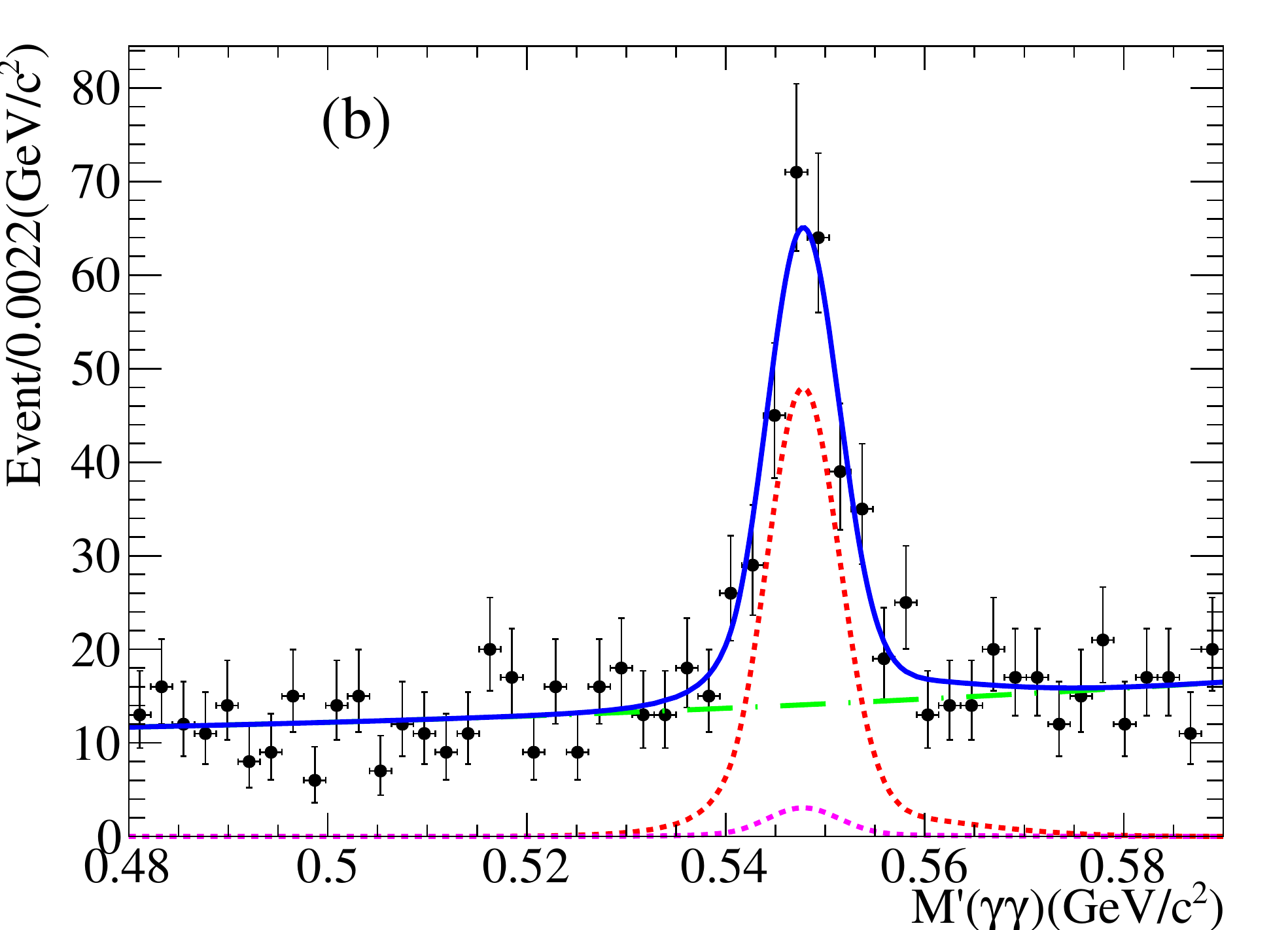}
\end{minipage}
\begin{minipage}{0.3\textwidth}
\centering
\includegraphics[width=\textwidth]{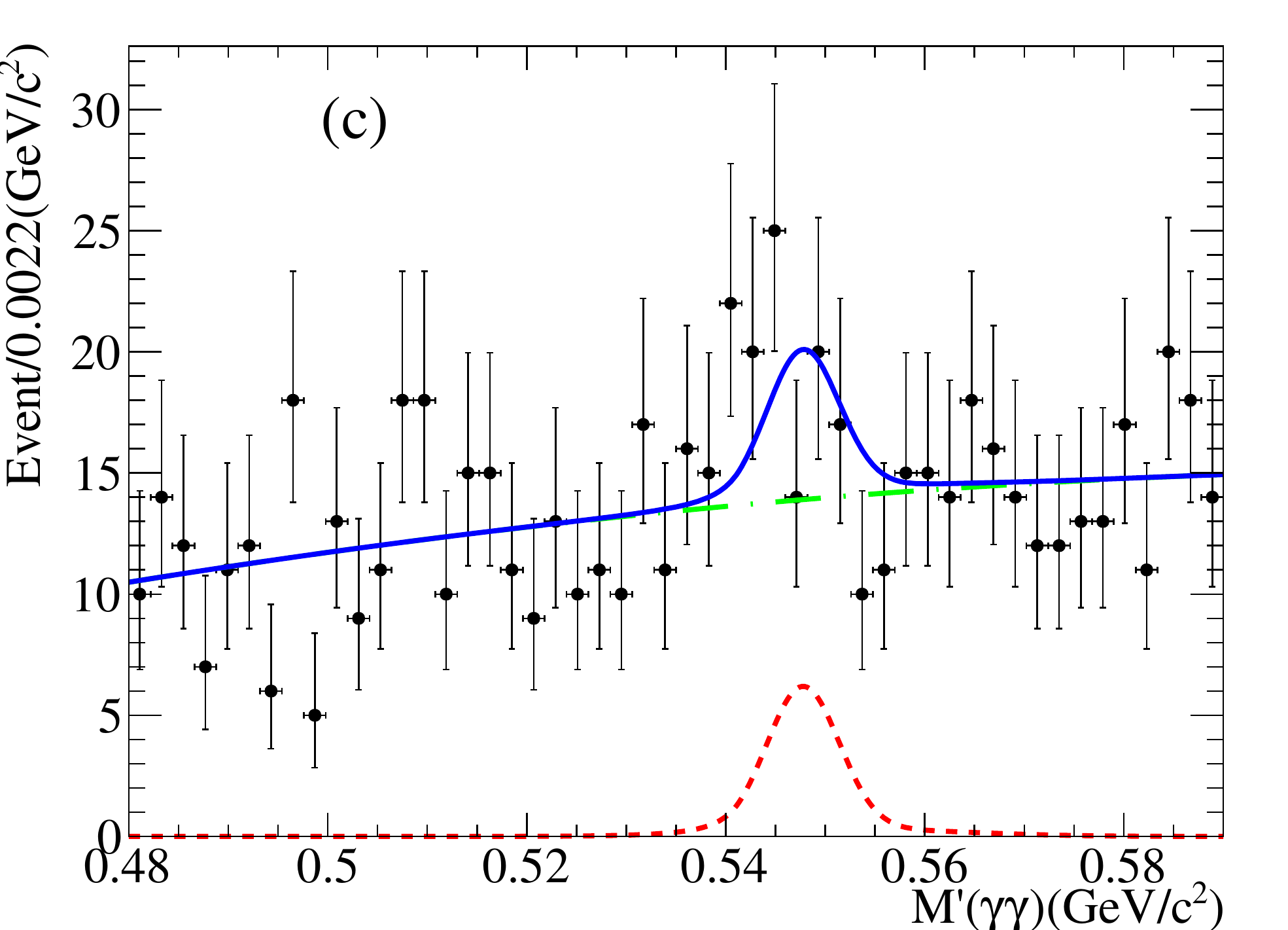}
\end{minipage}}
\caption{ Distribution of $M(\mu\mu)$ from data (a). The signal region is indicated 
by the two solid red arrows, while the sideband regions by the two dashed blue 
arrows. Distributions of $M'(\gamma \gamma)$ in the $J/\psi$ signal region (b) 
and sideband regions (c). The points with error bars are data, the blue solid 
curves represent the fit results, the red dashed curves represent signals components,
and the green dot-dashed curves represent background components.}
\label{figure2}
\ecl
\end{figure*}

The number of signal events 
is  obtained by an unbinned maximum-likelihood fit to the distributions of 
$M'(\gamma \gamma)$ in the $J/\psi$ signal region and sideband regions,  
with the $\eta$ signal line-shape shared for both regions. The $\jpsi$ 
signal region is defined as $M(\mu\mu)\in(3.06, 3.15)~\gevcc$ and the 
sideband regions as $M(\mu\mu)\in(2.9, 3.0)~\gevcc$ and $(3.2, 3.3)~\gevcc$.
The $\eta$ signal is described by the sum of a Crystal Ball function~\cite{Gaiser:1985ix} and a Gaussian function, while the combinatorial background is described by a second-order polynomial function.
The number of $\eta$ events in the sideband regions is multiplied by a scale factor $f$ and subtracted from
the number of $\eta$ event in the signal region, to give  the signal yield.
The scale factor 
$f=0.49$ is the ratio of non-$\jpsi$ events in the $\jpsi$ signal region and 
sideband regions, determined by a fit to the $M(\mu\mu)$ distribution. 
In the fit, the $\jpsi$ signal is described by the shape extracted 
from the signal MC simulation and the combinatorial background is described by 
a third-order Chebychev polynomial function. Figure~\ref{figure2} shows 
the distributions and fit results in $M(\mu\mu)$ and $M'(\gamma \gamma)$.
The observed signal yield 
is determined to be $ N^{\rm obs}=222\pm22$, where the uncertainty is statistical only. 

The Born cross section is determined by
\beq
{\sigma^{B}(e^+e^-\rightarrow\eta J/\psi)}= \frac{N^ {\rm obs}}{\mathcal{L}\cdot(1+\delta^{\rm ISR})\cdot(1+\delta^{\rm VP})\cdot\varepsilon\cdot{\cal{B}}r} \,,
\label{eq:born}
\eeq
where $\mathcal{L}$ is the integrated luminosity, $(1+\delta^ {\rm ISR})$ is the 
ISR correction factor~\cite{Kuraev:1985hb}, $(1+\delta^{\rm VP})$ is the vacuum
polarization factor taken from 
QED calculation~\cite{WorkingGrouponRadiativeCorrections:2010bjp}, ${\cal{B}}r$ 
is the product of the branching fractions of the subsequent decays of intermediate
states as given by the PDG~\cite{ParticleDataGroup:2020ssz}, and $\varepsilon$ 
is the detection efficiency. The ISR correction
factor is obtained by an iterative method~\cite{Sun:2020ehv}, in which the dressed cross
section
($\sigma^{D}(\sqrt{s})$=$\sigma^{B}(\sqrt{s})\cdot(1+\delta^{\rm
  VP})$) of $\EE\to\eta\jpsi$ measured in this study and previously
with c.m.\ energies from $\sqrt{s}=3.81$ to $4.60~\gev$~\cite{BESIII:2020bgb} are used as 
input. Table~\ref{table5} shows the measured Born cross section at 
$\sqrt{s}=3.773~\gev$ and the values of the other parameters in Eq.~\ref{eq:born}.  

\begin{table*}[htbp]
\caption{ The values of the integrated luminosity $\mathcal{L}$, the detection
efficiency $\varepsilon$, the product of radiative correction factor and vacuum
polarization factor $(1+\delta^{{\rm ISR}})\cdot(1+\delta^{VP})$, and the obtained 
Born cross section of $e^+e^-\rightarrow\eta J/\psi$ at $\sqrt{s}=3.773~\gev$. 
The uncertainties on the efficiency and cross section are statistical only.}
\begin{tabular}{ccccccc}\hline\hline
\multicolumn{1}{c}{\multirow {1}{*}{ $\mathcal L$ (pb$^{-1}$) }}  & \multicolumn{1}{c}{ $\varepsilon(\%) $ }  & \multicolumn{1}{c}{ $(1+\delta^{{\rm ISR}})\cdot(1+\delta^{VP})$ } & \multicolumn{1}{c}{ ${\cal{B}}(J/ \psi\rightarrow\mu^+\mu^-)(\%)$ } & \multicolumn{1}{c}{ ${\cal{B}}(\eta\rightarrow\gamma\gamma)(\%)$ } &  \multicolumn{1}{c}{ $N^{{\rm obs}}$ } & \multicolumn{1}{c}{ $\sigma^B$(pb) } \\
\hline
 $2931\pm15$ &  $45.4\pm0.1$  & $0.80$ & $5.96\pm0.03$  & $39.4\pm0.2$ & $222\pm22$  &  $8.89\pm0.88$ \\\hline\hline
\end{tabular}
\label{table5}
\end{table*}

The following sources of the systematic uncertainty are considered in the  cross-section measurement. The uncertainty on the integrated luminosity is 
$0.5\%$~\cite{BESIII:2018dpx}. The uncertainty associated with the reconstruction efficiency 
of an individual lepton or photon is
$1.0\%$~\cite{BESIII:2013ris,BESIII:2012fdg,BESIII:2015wyx}, giving 2\% for each pair of particles. 
An uncertainty of 1\% associated with the $J/\psi$ mass window requirement is assigned by 
comparing the $J/\psi$ mass resolution between data and MC simulation, and taking the  
difference in the selection efficiency. The 
helix parameters of the charged tracks are corrected in simulation to improve the agreement 
of $\chi^{2}_{\rm 4C}$ between data and MC simulation~\cite{BESIII:2012mpj}; 
the systematic uncertainty from the kinematic fit is estimated by removing the
correction and taking the 0.6\% difference in the detection efficiency as the
uncertainty. The systematic uncertainty from the ISR correction factor associated 
with the input cross section line-shape is  
estimated by sampling the parameters of the dressed cross-section 
line-shape using a multidimensional Gaussian function. The resultant distribution of $(1+\delta^{\rm ISR})$ values  
is fitted with a Gaussian function and the standard deviation of 0.5\%  is assigned as the
systematic uncertainty. In addition, the $\psi(3770)$ and $\psi(4040)$ resonance  parameters are varied within their uncertainties and the parametrization of the continuum  is considered by including a $1/s^n$ term, giving a 2.3\% uncertainty.  
The uncertainties on the quoted branching fractions of the decays of the intermediate states  
are taken from the PDG~\cite{ParticleDataGroup:2020ssz}, and lead to a 0.8\% uncertainty on the cross section. 
To determine the uncertainty associated with the fit procedure, we perform 
alternative fits by varying the resolution of the signal shape, the order of the
polynomial background shape, the normalization 
factor, and the fit range, individually, and use the difference in results to assign a 2.0\% uncertainty. The total systematic uncertainty is 
obtained to be 4.7\% adding all the individual items in quadrature, where the dominant contribution is from the background shape. 
The systematic uncertainty from each source is given in the Supplemental Material~\cite{Supplemental Material}.

\begin{figure}
\centering
\begin{overpic}
[width=7cm] {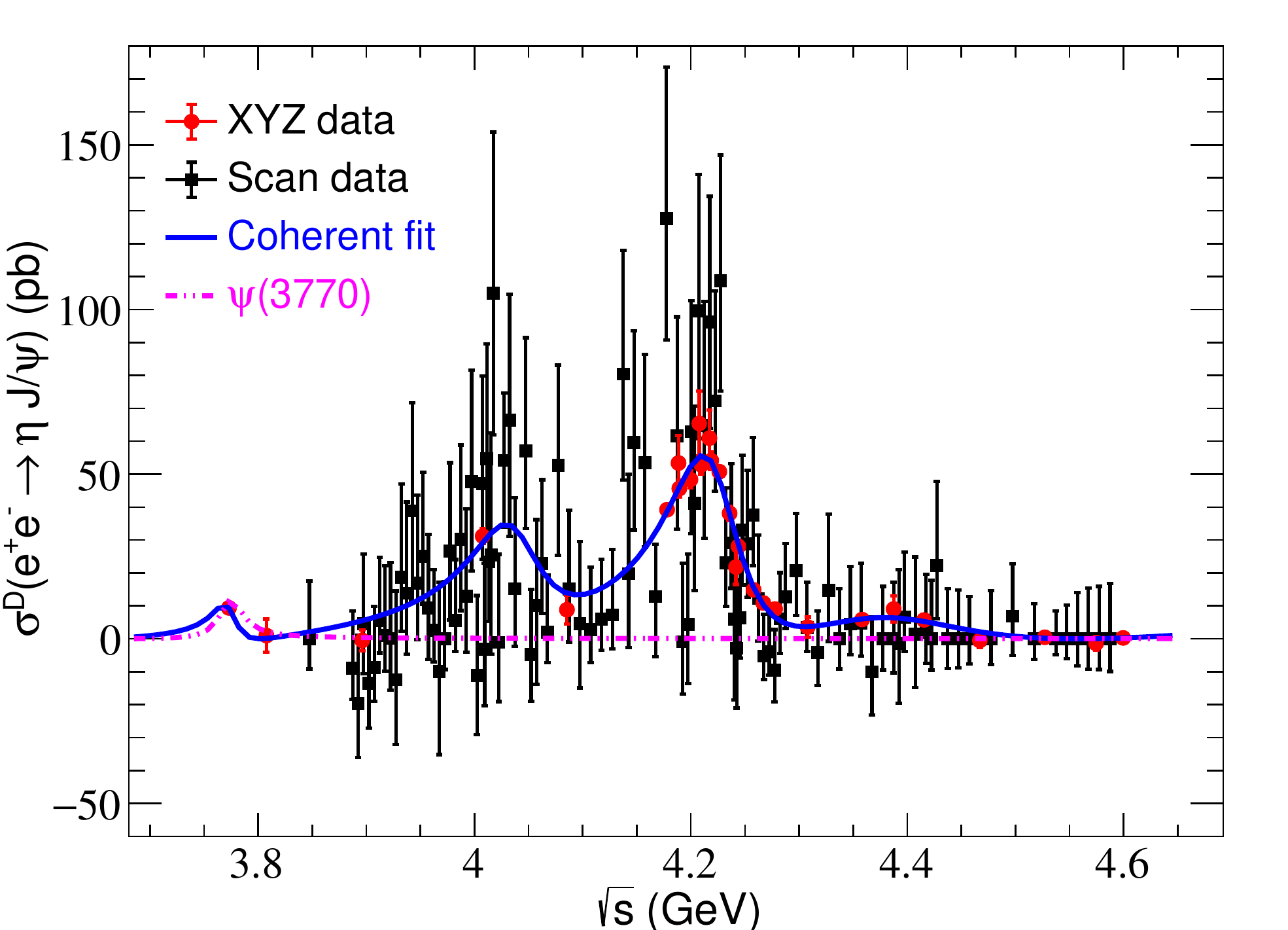}
\put(59, 40)
{\includegraphics[scale=0.12]{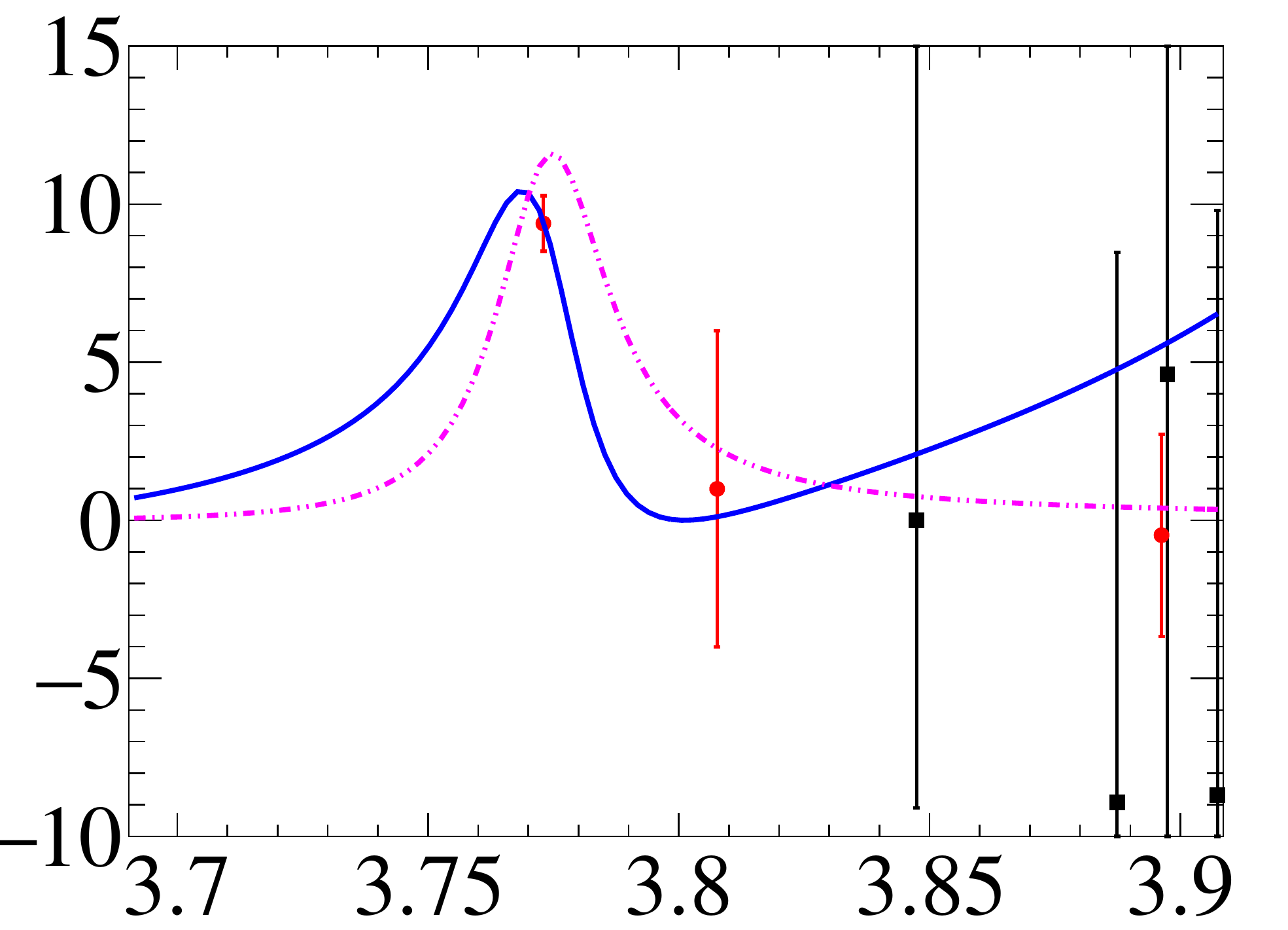}} 
\end{overpic}

\begin{overpic}
 [width=7cm]{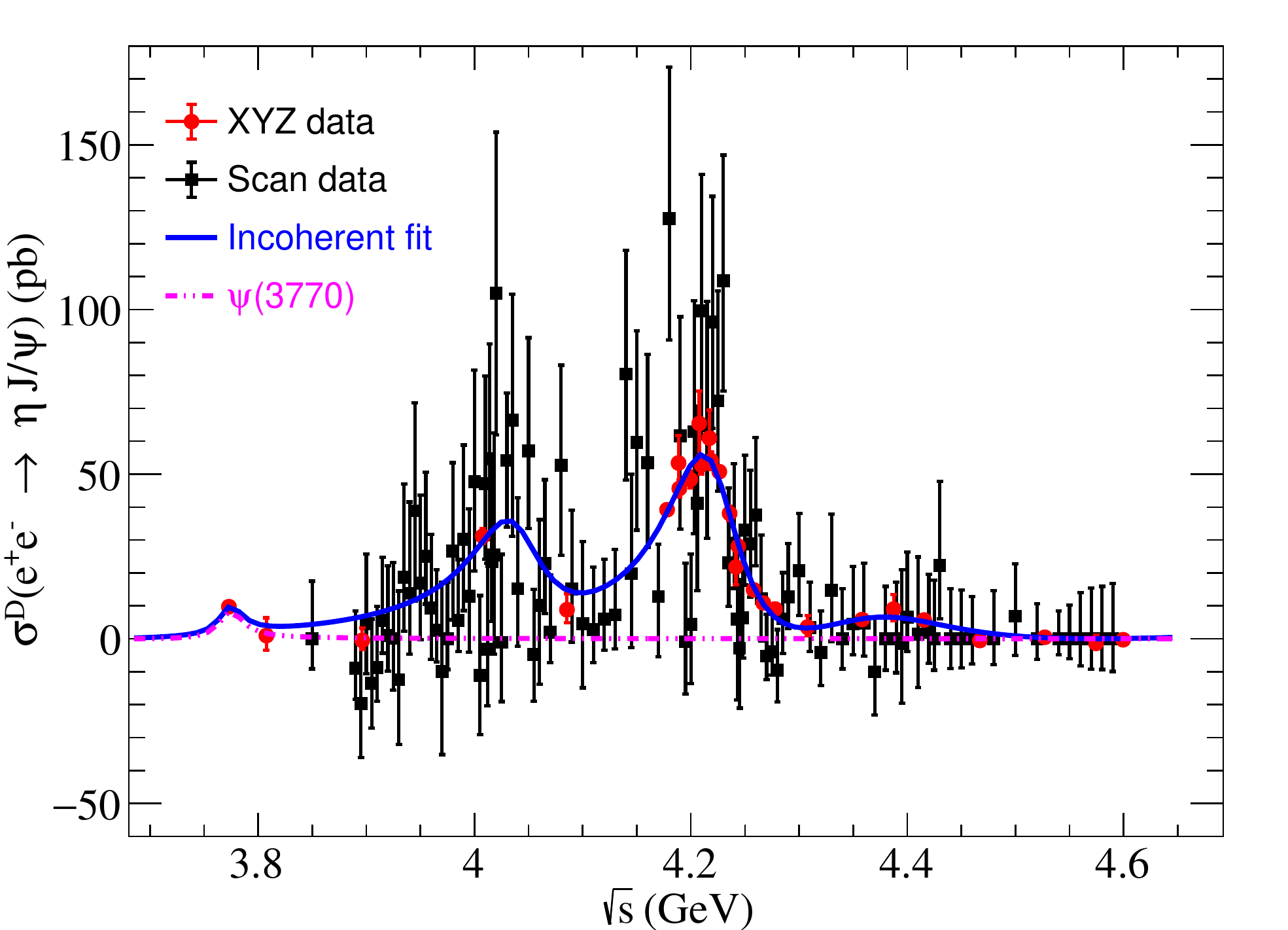}
 \put(59, 40) 
{\includegraphics[scale=0.12]{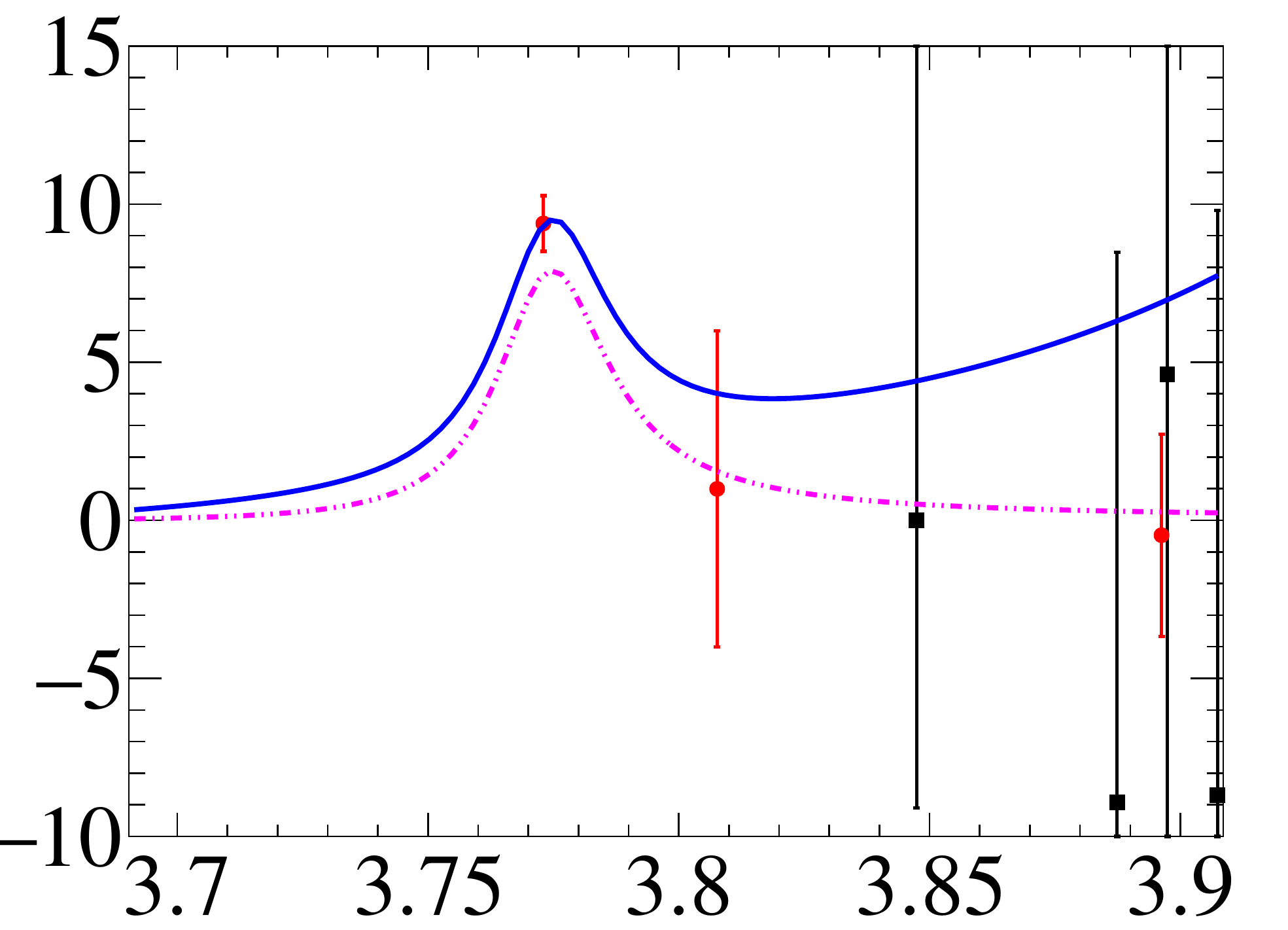}}
 \end{overpic}
\caption{(Top) Coherent and (bottom) incoherent fits to the dressed cross section line-shape of $\EE\to\eta\jpsi$. The points with 
error bars are data and the solid curves are the best fit results. The insert is the zoomed distribution 
in the $\psipp$ mass region.}
\label{figure3}
\end{figure}

The branching fraction of $\psipp\to\eta\jpsi$ is determined from a maximum 
likelihood fit to the dressed cross section of $\EE\to\eta\jpsi$ from 
$\sqrt{s}=3.773$ to $4.6~\gev$. The likelihood is constructed taking the 
fluctuations of the number of signal and background events into
account~\cite{BESIII:2016adj}.
Two scenarios are used to describe the dressed cross section line-shape, with two different treatments of the $\psi(3770)$ resonant decay amplitude: one in which the $\psi(3770)$ contribution is coherent
\begin{eqnarray}
\sigma_{\rm co.} & =|C\cdot\sqrt{\Phi(s)}+e^{i\phi_1}\bw_{\psipp}+e^{i\phi_2}\bw_{\psi(4040)} \nonumber  \\
                              & +e^{i\phi_3}\bw_{Y(4230)}+e^{i\phi_4}\bw_{Y(4390)}|^2 \, ,
\end{eqnarray}
and the other where it is incoherent
\begin{eqnarray}
\sigma_{\rm inco.} & = |\bw_{\psi(3770)}|^2 +|C\cdot\sqrt{\Phi(s)}+e^{i\phi_2}\bw_{\psi(4040)} \nonumber \\
                                & +e^{i\phi_3}\bw_{Y(4230)}+e^{i\phi_4}\bw_{Y(4390)}|^2 \, .
\end{eqnarray}
Here $\Phi(s)=q^{3}/s$ is the $P$-wave PHSP factor used to parameterize the continuum term, with $q$ being the 
$\eta$ momentum in the $\EE$ c.m. frame,
$\bw=\frac{\sqrt{12\pi \br\Gamma_{ee}\Gamma}}{s-M^{2}+iM \Gamma}\sqrt{\frac{\Phi(s)}{\Phi(M^2)}}$ 
is the Breit-Wigner function, $\phi$ is the relative phase between the resonant
decay and the PHSP term, and $C$ is a real parameter. In the Breit-Wigner formula, $M$, $\Gamma$, and $\Gamma_{ee}$ and 
$\br$ are the mass, the total width, the electronic width (whose definition includes vacuum polarization effects), and the branching
fraction to $\eta\jpsi$ of the resonance. The mass and total width of $\psipp$ 
and $\psi(4040)$, and the electronic width of $\psipp$ are fixed to the PDG
values~\cite{ParticleDataGroup:2020ssz}, while $\br$ and the
parameters of the other resonances are free parameters to be
determined by the fit. Only the statistical uncertainty of the 
dressed cross section is considered in the fit. There are four solutions from the coherent 
fit and one solution from the incoherent fit. Figure~\ref{figure3} shows the cross-section measurements plotted against $\sqrt{s}$, with the
fit results superimposed.  The result for the coherent fit is degenerate for the four solutions. 
  The fit qualities estimated by a $\chi^2$-test approach
are $\chi^2/n.d.f. = 102.4/119$ for the coherent fit and $106.9/120$ for the incoherent fit, 
where $n.d.f$ is the number of degrees of freedom. The statistical significance 
of the $\psi(3770)\to\eta\jpsi$ decay in the coherent (incoherent) fit is estimated to be $7.4\sigma$ ($7.6\sigma$), calculated by the change of the likelihood values with and without 
the $\psipp$ resonance contribution included, and taking the change in the number of degrees 
of freedom into account~\cite{Wilks:1938dza}. The branching fractions from the fits are summarized in Table~\ref{table10}. The results of the other resonant parameters are consistent with those found in the earlier study~\cite{BESIII:2020bgb}.   The statistical uncertainty of the coherent fit  is large due to the lack of 
data points around the $\psipp$ peak, which  leads to a poor constraint on the 
relative phase $\phi_1$.

\begin{table}[htbp]
\caption{The branching fractions of $\psipp\to\eta\jpsi$ determined from the 
coherent and incoherent fits.}
 \bcl
\begin{tabular}{cccc}\hline\hline
\multicolumn{2}{c}{\multirow{2}{*}{Fit scenario}} &  $\br(\psi(3770)\rightarrow\eta J/\psi)$ & \multirow{2}{*}{$\phi_1$(rad)} \\
 &&$(\times 10^{-4})$ & \\
\hline
\multirow{4}{*}{~~Coherent fit~~}  &Solution1& $11.6 \pm 6.1 \pm 1.0$ &4.0$\pm$0.5\\
                                                      &        Solution2 & $12.0 \pm 6.3 \pm 1.1$ &4.3$\pm$0.5\\
                                                      &        Solution3 & $11.6 \pm 6.1 \pm 1.0$ &3.8$\pm$0.5\\
                                                      &        Solution4 & $11.9 \pm 6.3 \pm 1.1$ &4.1$\pm$0.5\\\hline
{\hspace*{0.2cm} Incoherent fit} & & \phantom{0}$7.9 \pm 1.0 \pm 0.7 $ &- \\\hline\hline
\end{tabular}
\label{table10}
\ecl
\end{table}

There are several sources of potential systematic bias in the  branching-fraction measurement.
By way of example we quote the uncertainties for solution 1 of the coherent fit.
 The uncertainty of the c.m. energy is $0.8~\mev$~\cite{BESIII:2016bnd} for all data samples; this uncertainty is propagated to the branching-fraction measurement to give a
relative uncertainty of 0.5\%. The uncertainty from the energy spread is 0.1\%, which is estimated by 
convolving the fit formula with a Gaussian function with a standard deviation of 
$1.4~\mev$, which is the measured value of the spread~\cite{BESIII:2016adj, Sect:2011}. The uncertainty arising from the
$\psi(3770)$ and $\psi(4040)$ resonant parameters is studied by varying the 
parameters within their uncertainties, which leads to an effect of 8.1\%, where the dominant contribution (6.9\%) is from the partial width to di-electrons. The uncertainty of the parameterization of 
the continuum term is assigned to be 0.8\% by changing the $1/s$ dependence to $1/s^n$, where $n$ is
a free parameter. The uncertainty from the dressed cross-section measurement has two contributions:
the one of 1.7\% is uncorrelated among the c.m.\ energy points and is included in the fit to the dressed cross section; the other of 3.0\%  is common to all data points and is directly propagated to the $\br$ measurement. 
The total systematic uncertainty is $8.9\%$, and the individual contributions are listed in the Supplemental Material~\cite{Supplemental Material}.

In summary, the Born cross section of $ e^+e^-\rightarrow\eta J/\psi $ at 
$\sqrt s=3.773~\gev$ is measured to be $\sigma^B(e^+e^-\rightarrow\eta J/\psi) = (8.89\pm0.88_{\rm stat}\pm0.42_{\rm sys})~\pb$. 
The decay $\psipp\to\eta\jpsi$ is observed for the first time with a 
statistical significance of $7.4\sigma$. The branching fraction of 
$\psipp\to\eta\jpsi$ is determined from the fit 
to the dressed cross section line-shape of $\EE\to\eta\jpsi$ in the range of $\sqrt{s}=3.773$ 
to $4.60~\gev$ including the decays of the $\psipp$, $\psi(4040)$, 
$Y(4230)$ and $Y(4390)$ resonances as well as the PHSP term.
When the interference of the decay of the $\psi(3770)$ with the other processes is neglected,
the  branching fraction 
is determined to be $(7.9 \pm 1.0_{\rm stat} \pm 0.7_{\rm sys} )\times10^{-4}$,
which is close to the result of  CLEO~\cite{CLEO:2005zky} but with twice the precision. 
When interference is considered, four solutions
are obtained with branching fractions varying between 
$(11.6 \pm 6.1_{\rm stat} \pm 1.0_{\rm sys})\times10^{-4}$ and $(12.0 \pm 6.1_{\rm stat} \pm 1.1_{\rm sys})\times10^{-4}$.
 These results would benefit, as essential inputs, to the calculations of charmonia decaying into light vector pseudo-scalar (VP) states~\cite{Zhang:2009kr,Li:2013zcr} and hadronic transitions of highly excited charmonium(-like) states~\cite{Anwar:2016mxo}. And we notice the measured branching fractions are close to the predicted value of Ref.~\cite{Voloshin:2005sd}
and hint at a possible tetra-quark component in the $\psi(3770)$ resonance. A finer scan around the $\psi(3770)$ is desirable to reduce the uncertainties in the future.

\section{ACKNOWLEDGMENTS}
The BESIII collaboration thanks the staff of BEPCII and the IHEP computing center for their strong 
support. This work is supported in part by National Key  Research and Development Program of China 
under Contracts Nos. 2020YFA0406300, 2020YFA0406400; National Natural Science Foundation of China 
(NSFC) under Contracts Nos. 11625523, 11635010, 11735014, 11822506, 11835012, 11935015, 11935016, 
11935018, 11961141012, 12022510, 12025502, 12035009, 12035013, 12061131003; the Chinese Academy of 
Sciences (CAS) Large-Scale Scientific Facility Program; Joint Large-Scale Scientific Facility Funds 
of the NSFC and CAS under Contracts Nos. U1732263, U1832207, U2032108; CAS Key Research Program 
of Frontier Sciences under Contract No. QYZDJ-SSW-SLH040; the CAS Center for Excellence in Particle Physics 
(CCEPP); 100 Talents Program of CAS; INPAC and 
Shanghai Key Laboratory for Particle Physics and Cosmology; ERC under Contract No. 758462; European 
Union Horizon 2020 research and innovation programme under Contract No. Marie Sklodowska-Curie grant 
agreement No 894790; German Research Foundation DFG under Contracts Nos. 443159800, Collaborative 
Research Center CRC 1044, FOR 2359, GRK 2149; Istituto Nazionale di Fisica Nucleare, Italy; 
Ministry of Development of Turkey under Contract No. DPT2006K-120470; National Science and Technology 
fund; Olle Engkvist Foundation under Contract No. 200-0605; STFC (United Kingdom); The Knut and Alice 
Wallenberg Foundation (Sweden) under Contract No. 2016.0157; The Royal Society, UK under Contracts 
Nos. DH140054, DH160214; The Swedish Research Council; U. S. Department of Energy under Contracts 
Nos. DE-FG02-05ER41374, DE-SC-0012069.

\end{document}